\documentclass{WileyMSP-template}

\usepackage{amsmath}
\usepackage[noadjust, nocompress]{cite}
\usepackage{xcolor}

\let\citesuper=\cite
\renewcommand{\cite}[1]{\textsuperscript{{\citesuper{#1}}}}

\begin{document}


\title{Extremely large nonlinear response in crystalline quartz at THz frequencies}

\maketitle


\author{Soheil Zibod*}
\author{Payman Rasekh}
\author{Murat Yildrim}
\author{Wei Cui}
\author{Ravi Bhardwaj}
\author{Jean-Michel M\'enard}
\author{Robert W. Boyd}
\author{Ksenia Dolgaleva}



\begin{affiliations}
S.~Zibod, P.~Rasekh\\
School of Electrical Engineering and Computer Science, University of Ottawa, Ottawa, Ontario K1N 6N5, Canada\\
Email Address: szibo043@uottawa.ca\\

M.~Yildirim, W.~Cui, R.~Bhardwaj, J.-M.~M\'enard\\
Department of Physics, University of Ottawa, Ottawa, Ontario K1N 6N5, Canada\\

R.~W.\ Boyd\\
Department of Physics, University of Ottawa, Ottawa, Ontario K1N 6N5, Canada; Institute of Optics, University of Rochester, Rochester, New York 14627,United States\\

K.~Dolgaleva\\
School of Electrical Engineering and
Computer Science, University of Ottawa, Ottawa, Ontario K1N 6N5, Canada; Department of Physics, University of Ottawa, Ottawa, Ontario K1N 6N5, Canada

\end{affiliations}



\justifying

\begin{abstract}
We report on the first experimental observation of a very strong nonlinear response in crystalline quartz in the terahertz (THz) frequency region through THz time-domain spectroscopy (THz-TDS). Theoretical modelling is presented and predicts a Kerr coefficient $n_2$ equal to \( 5.17 \times 10^{-14} \  \mathrm{m^2\ W^{-1}} \). The time-domain analysis of the measured data shows that with increasing of the THz peak amplitude, the pulse experiences a larger time delay in the sample. As the THz amplitude increases to values higher than 110~$\mathrm{kV\ cm^{-1}}$, the growth rate of the delay decreases, indicating a saturation process. The value of the nonlinear refractive index calculated through the frequency-response analysis is estimated to be on the order of \(10^{-13} \  \mathrm{m^2\ W^{-1}} \),  which is several orders of magnitude larger than typical values of the nonlinear refractive index of solids in the visible region. Furthermore, a negative fifth-order susceptibility on the order of \(10^{-30} \ \mathrm{m^4\ V^{-4}}\) is measured. 


\end{abstract}


\section{Introduction}
Terahertz (THz) radiation, defined as a region of the electromagnetic spectrum between the microwaves and far-infrared, is gaining a growing importance in applications such as biomedical sensing,\cite{Peng2020,Lindley2021} security,\cite{Chen:07} spectroscopy and imaging,\cite{Jepsen2011} and communications.\cite{Kleine-Ostmann2011} Furthermore, THz time-domain spectroscopy (THz-TDS) systems are used for monitoring production processes,\cite{krumbholz2009monitoring} art conservation,\cite{Krugener2015} and characterization of materials.\cite{busch2014optical}

THz-TDS allows the simultaneous measurement of the magnitude and phase of the THz signal through the linear electro-optic effect, representing a suitable technique for measuring the complex refractive index of a material at the THz frequencies.\cite{Nahata:96}
The recent development of intense THz pulse generation techniques opens the door to studying nonlinear behavior of different materials in the THz region.\cite{Hebling:08} Nonlinear effects such as THz-induced impact ionisation and inter-valley scattering in semiconductors,\cite{Asmontas2020,Hoffmann2009,Lange2014,Tarekegne2015,Blanchard2012} THz high-harmonic generation by hot carriers,\cite{Schubert2014,Hafez2018,Chai2018,Gaal:2006} and THz-induced ferroelectricity and collective coherence control have been demonstrated.\cite{Nelson2009,Nelson2019}
A very large third-order nonlinearity has been reported for water vapor,\cite{Rasekh2021} where the stepwise multiphoton transitions in water molecules lead to a third-order susceptibility of $\chi^{(3)}=(0.4+6i)\times 10^2 \ \mathrm{m^2 \ V^{-2}}$. Extreme THz-induced Kerr effects have been
reported for  different liquids,\cite{Boyd2021,Kerr2009,Tcypkin2019,Francis2020} where the nonlinear refractive indices can be several orders of magnitude larger than their values in the optical regime. Moreover, THz-induced Kerr effects have been observed in amorphous chalcogenide glasses such as arsenic trisulfide and arsenic triselenide.\cite{Zalkovskij2013} Furthermore, it has been theoretically predicted that crystals can exhibit an extremely large nonlinear refractive index in the THz frequency range.\cite{Dolgaleva2015} Crystalline solids such as quartz are predicted to show THz nonlinear refractive indices that exceed the optical values by several orders of magnitude. However, there has not been any experimental demonstration reported to date.

Here we report on the experimental observation of very strong nonlinear interactions in crystalline quartz in the THz regime.
First, a theoretical model for the nonlinear refractive index of quartz at the THz frequencies  is  presented.\cite{Dolgaleva2015} This model relies on the classical anharmonic oscillator, where the nonlinear refractive index is given as a sum of contributions from different vibrational modes. As predicted by the model, the value of the nonlinear refractive index at the lower frequencies exceeds its typical values in the visible range by several orders of magnitude. Then, we perform nonlinear THz-TDS on a 1-mm-thick z-cut quartz sample. The time-domain analysis of the collected data demonstrates an increased delay, experienced by the pulse as it propagates through the sample, with increasing THz beam intensity. However, the growth rate of the delay decreases with the further intensity increase, revealing a phase saturation process. Further, the analysis in the Fourier domain shows an increase in the nonlinear phase and nonlinear absorption with the increase of the THz field intensity. At higher signal levels, however, the nonlinear phase grows with field intensity increase at a declining pace, whereas the nonlinear absorption tends to increase more rapidly. The data analysis revealed extremely large values of the nonlinear refractive index and fifth-order susceptibility, where the latter has a negative real part.

The manuscript is structured as follows: In Section 2, we describe a simple theoretical model for calculating the contributions from the vibrational modes to the nonlinear refractive index of crystalline quartz at THz frequencies, developed originally in Ref.~\cite{Dolgaleva2015}. We modify the model to include additional vibrational resonances, which helps us to achieve a better agreement with the experimental results. In Section 3, the experimental setup for the nonlinear THz-TDS is introduced. In section 4, we present the experimental results. In addition to the time-domain analysis, the frequency response of crystalline quartz is presented. Finally, the nonlinear parameters of the material are calculated and the result is compared with the theoretical model.

\begin{figure*}[!t]
\centering
    \includegraphics[height=8cm]{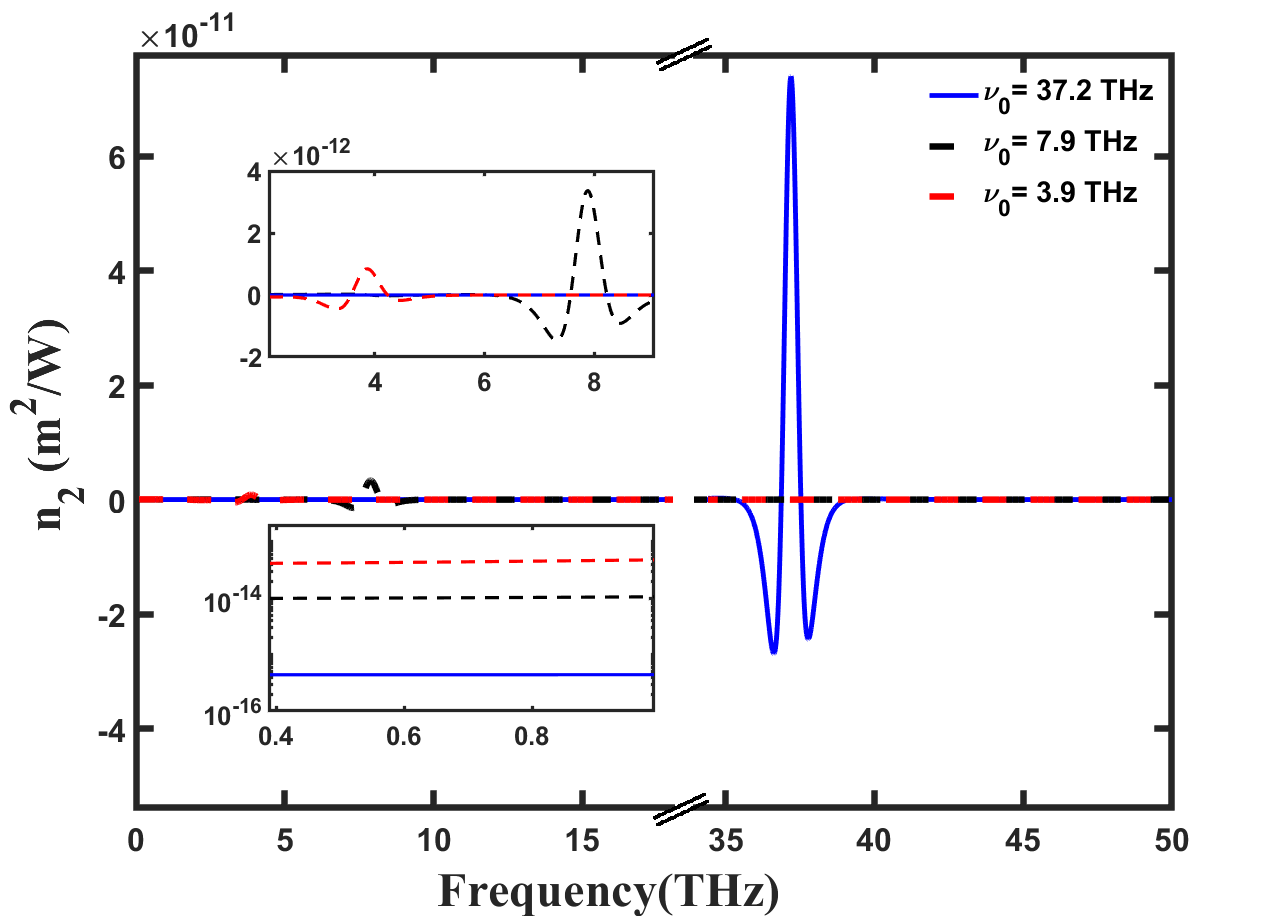}
\caption{\label{fig: sim} Theoretical modelling of the nonlinear refractive index caused by the dominant resonance at 37.2~THz (blue solid lines) and
the resonances at 7.9~THz (black dashed lines) and 3.9~THz (red dashed lines) in quartz. The top inset resolves the values of the nonlinear refractive indices at around the low-frequency resonances. The bottom inset shows the contribution from three resonances at the lower frequencies (1~THz and lower).  }

\end{figure*}
\section{Theory}
An extremely large refractive index has been theoretically predicted for quartz in the THz regime, where the nonlinear refractive index was predicted to be several orders of magnitude larger than its typical visible and near-infrared values.\cite{Dolgaleva2015} The model is based on the equation of motion of a classical anharmonic oscillator:
\begin{equation}
   \ddot{x} + 2\gamma \dot{x}+ \omega_0^2 x + a x^2 + b x^3=\alpha E,
   \label{equ_anharmonic}
\end{equation}
where $x$ is the ion displacement from the equilibrium position, $\gamma$ is damping factor, $\omega_0$ is the resonance frequency, $E$ is the applied field and $a$ and $b$ are the second- and third-order nonlinear coefficients, respectively. The parameter $\alpha$ on the right-hand side of Equation~(\ref{equ_anharmonic}) is determined as $\alpha = {q}/{m}$, where $q$ and $m$ are the effective electric coupling strength and effective reduced mass of the vibrational mode, respectively. After applying perturbation theory and performing some algebraic operations,\cite{Dolgaleva2015} the relationship between the complex nonlinear refractive index and the resonance parameters takes the form
\begin{equation}
   \tilde{\bar{n_2}}=\frac{\pi q N}{\tilde{n_0}} \frac{\alpha^3}{(\omega_0^2-\omega^2-2\gamma i \omega)^4} \times \left[ 2a^2 \frac{3\omega_0^2-8\omega^2-8\gamma i \omega}{\omega_0^2 (\omega_0^2-4\omega^2-4\gamma i \omega)} + 3b \right ].
   \label{equ_n}
\end{equation}
Here $N$ is the atomic density, and $\tilde{n_0}$ is the linear refractive index. Among the two terms, the term related to the second-order nonlinearity is two orders of magnitude larger than the contribution coming from the third-order nonlinear coefficient, $b$ (see details in Supporting Information). This can be attributed to the fact that the cascaded processes are usually much stronger than the direct higher-order processes. The nonlinear coefficient $a$ is related to the known parameters of the crystal through
\begin{equation}
 a=-\frac{a_1 m \omega_0^4}{k_B} \alpha_T, 
 \label{equ_a}
\end{equation}
where $a_1$ is the lattice constant, $k_B$ is Boltzmann constant and $\alpha_T$ is the thermal expansion coefficient. With the assumption of a single dominant vibrational mode at 37.2~THz and ignoring the much weaker resonances at lower frequencies, one can evaluate the nonlinear refractive index of crystalline quartz at very low frequencies to be $\bar{n_{2}}^{\omega \ll \omega_0} \approx 2.21 \times 10^{-9} $~esu or, equivalently, $4.42 \times  10^{-16} $~$\mathrm{m^2\ W^{-1}}$.\cite{Dolgaleva2015}  \\

However, the strong vibrational resonance at 37.2~THz is not the only resonance contributing to the vibrational $n_2$. There are several other resonances at lower frequencies, among which are the ones at 3.9~THz and 7.9~THz.\cite{Davies2018} To take the contribution of these additional resonances into consideration, we modify Equation~(\ref{equ_n}) into the form 
\begin{equation}
   \tilde{\bar{n_2}}=\frac{\pi q N \alpha^3}{\tilde{n_0}}\sum_{j=1}^3 \frac{1}{(\omega_{0,j}^2-\omega^2-2\gamma i \omega)^4} \times \left[ 2a_j^2 \frac{3\omega_{0,j}^2-8\omega^2-8\gamma i \omega}{\omega_{0,j}^2 (\omega_{0,j}^2-4\omega^2-4\gamma i \omega)} + 3b_j \right ].
   \label{equ_sum}
\end{equation}
In this modified equation, the nonlinear refractive index is now given as a sum of the contributions from the three resonances: the strong resonance at 37.2~THz, and the weaker resonances at 3.9~THz and 7.9~THz. Equation~(\ref{equ_a}) shows that the second-order nonlinear coefficient is proportional to $\omega_0^4$ (see details in Supporting Information), meaning that the resonant value of the nonlinear coefficient for the dominant resonance is, for instance, approximately 500 times larger than that for the resonance at 7.9~THz. Figure \ref{fig: sim} shows the dispersion of the nonlinear refractive index caused by the dominant resonance and the resonances at 7.9~THz and 3.9~THz. The top inset of Figure \ref{fig: sim} resolves the value of $n_2$ at around low-frequency resonances. The comparison indicates that the nonlinear refractive index at the dominant resonance at 37.2~THz is approximately 20 times larger than the one at the stronger resonance of the two lower-frequency resonances.

Substituting Equation~(\ref{equ_a}) into Equation~(\ref{equ_sum}), we can see that the contributions of different resonances to the nonlinear refractive index at very low frequencies are proportional to $1/\omega_0^2$. Consequently, at much lower frequencies (1~THz and below), the contributions from the resonances at 7.9~THz and 3.9~THz are approximately 20 times larger and 100 times larger than the one at 37.2~THz, respectively, as shown in the bottom inset of Figure \ref{fig: sim}. The contributions of the three resonances to the nonlinear refractive index are listed in Table \ref{tab:1}. 
The table clearly demonstrates that the dominant contributions to the nonlinear refractive index at very low frequencies are from the resonances at 7.9~THz and 3.9~THz: $n_{2}^{\omega \ll \omega_0}$ $\approx$ $5.17 \times 10^{-14}  \ \mathrm{m^2\ W^{-1}} $.
\begin{table}[h!]
\centering
 \caption{\label{tab:1}The contribution to the nonlinear refractive index }
  \begin{tabular}[htbp]{@{}lll@{}}
    \hline
    $\omega_0$ (THz) & $n_{2}^{\omega \ll \omega_0}$ ($\mathrm{m^2\ W^{-1}}$) & $n_2^{\omega \approx \omega_0}$  ($\mathrm{m^2\ W^{-1}}$) \\
    \hline
    3.9   & $4.17 \times 10^{-14}$   & $8.38 \times 10^{-13}$  \\
    7.9   & $9.98 \times 10^{-15}$   & $3.35 \times 10^{-12}$  \\
    37.2  & $4.42 \times 10^{-16}$   & $7.40\times 10^{-11}$   \\
    \hline
  \end{tabular}
  \end{table}

\section{Experiment}
\begin{figure*}[!t]
\centering
        
    \includegraphics[width=12cm,height=8cm]{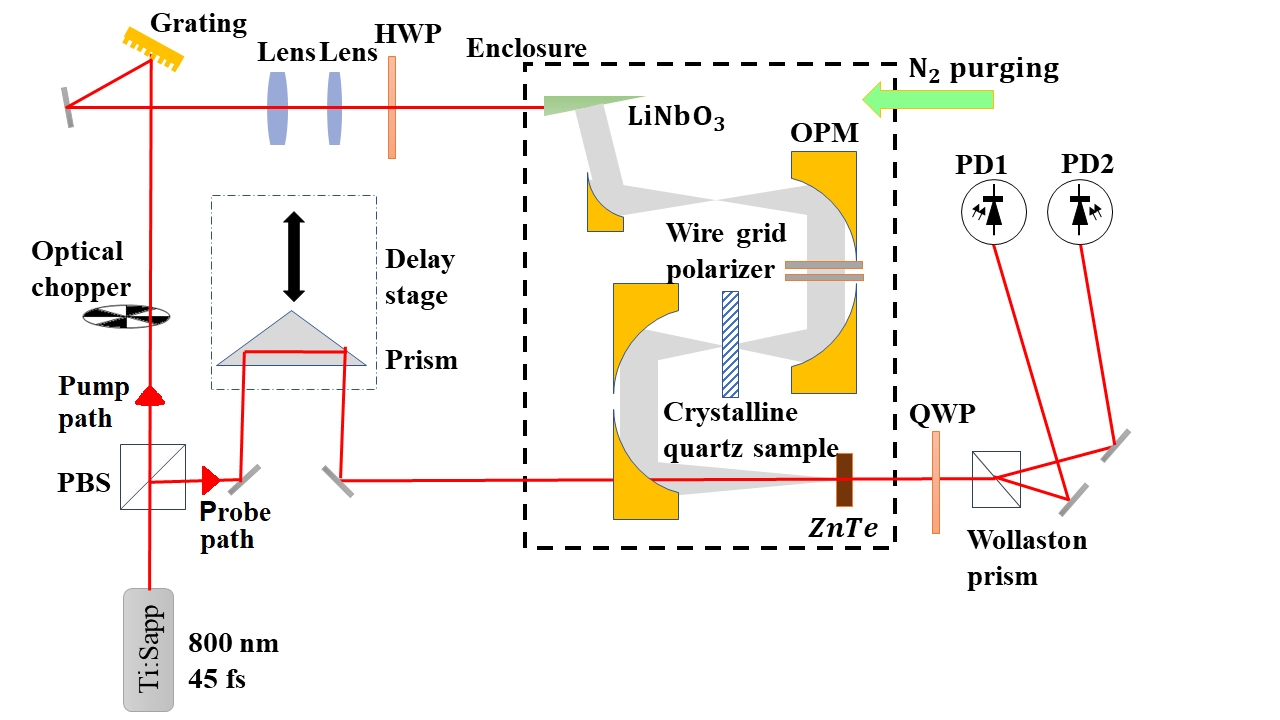}
\caption{\label{fig:1} The schematic of THz-TDS experimental setup. The 800-nm beam is split into pump and probe paths. The phase-matching condition required for the generation of intense THz field in $\mathrm{LiNbO_3}$ crystal is achieved through the pulse-front tilting technique. OPM: Off-axis parabolic mirror; HWP: Half-wave plate; QWP: Quarter-wave plate; PD: Photodetector. }
\end{figure*}

The intense THz radiation is generated in an optical rectification process in lithium niobate ($\mathrm{LiNbO_3}$), where the pulse-front tilting technique is used to make the process phase-matched and efficient.\cite{Hebling:08} The setup schematic is depicted in Figure~\ref{fig:1}. The beam, coming from a 800-nm Ti:sapphire laser with a pulse duration of 45~fs and repetition rate of 1~kHz, is split into the pump and probe paths. In the pump path, the beam diffracts from a grating and, after passing through two cylindrical lenses, propagates through the generation crystal. The generated THz radiation is collimated and focused with several gold off-axis parabolic mirrors. A pair of wire-grid polarizers is also used to control the THz field amplitude during the measurements. 

In the probe path, the near-infrared (NIR) probe and THz beams co-propagate inside the 200-\(\mathrm{\mu }\)m-thick ZnTe detection crystal. A delay stage is also used to change the overlap time between the THz and probe beams, so that one can measure different points of the THz pulse. As the THz pulse propagates through the detection crystal, the refractive index experienced by the probe beam is modified through the linear electro-optic effect, resulting in a birefringence in the crystal. The phase difference induced by the birefringence is then converted into the beam's ellipticity via a quarter-wave plate. A Wollaston prism splits the beam into two components of which their intensity difference is proportional to the beam ellipticity. Finally, a pair of balanced photodetectors connected to the lock-in amplifier is used to detect the differential signal. The peak amplitude of the electric field is estimated to be 225~$\mathrm{kV\ cm^{-1}}$ at the focal position where we place the 1-mm $z$-cut quartz sample. To eliminate the water-vapor absorption, the part of the setup where the THz beam is generated and propagates is enclosed and purged with nitrogen. Different field amplitudes are obtained by rotating the first wire-grid polarizer and keeping the second one fixed.

\begin{figure*}[t]
\centering
   \begin{tabular}{ l r }
   \begin{tabular}{@{} l @{}}
    (a) \\[1ex]
    \includegraphics[width=12cm]{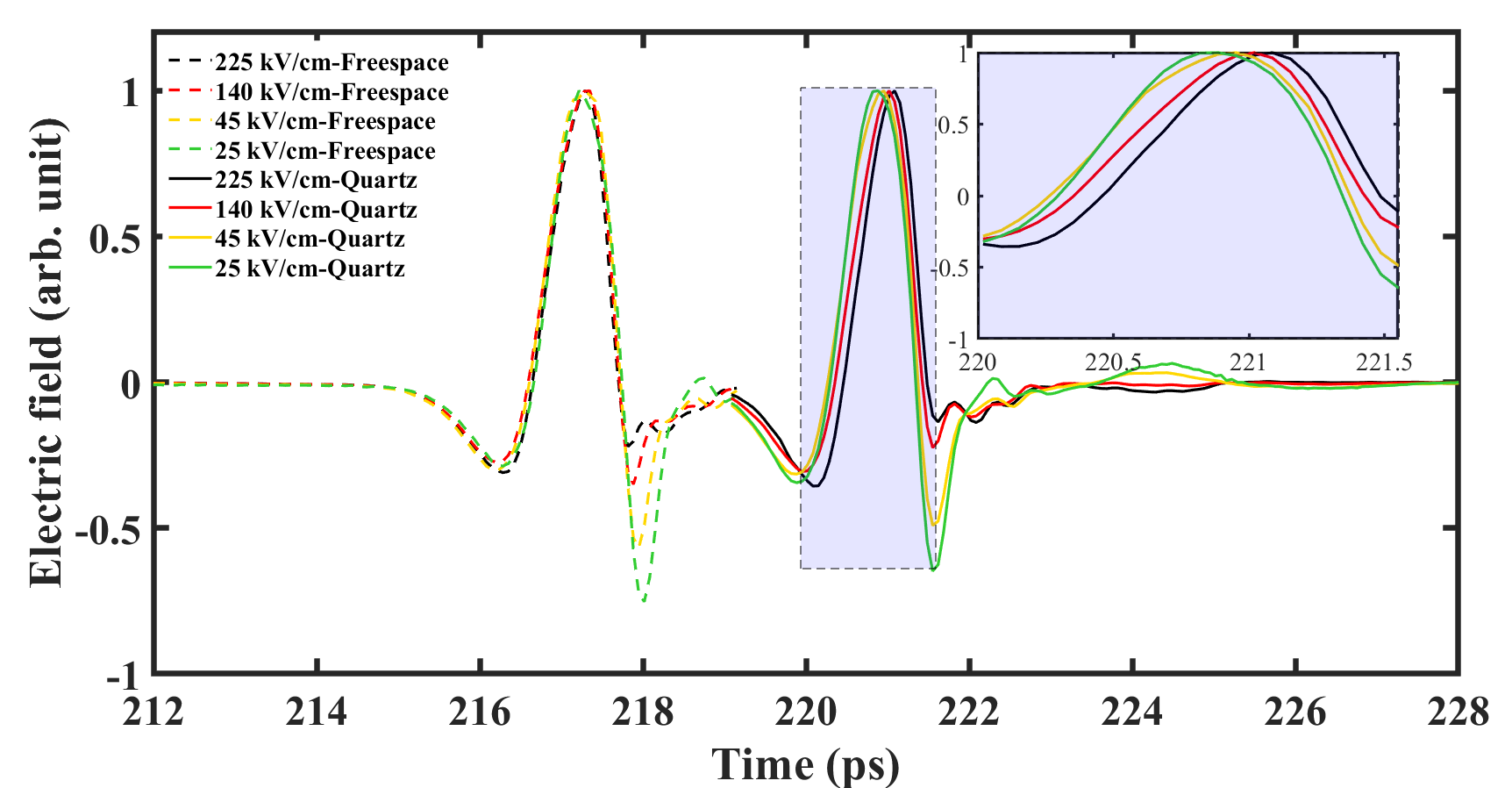}
  \end{tabular}
  &
  \begin{tabular}{@{} l @{}}
   (b)\\
    \includegraphics[width=7cm]{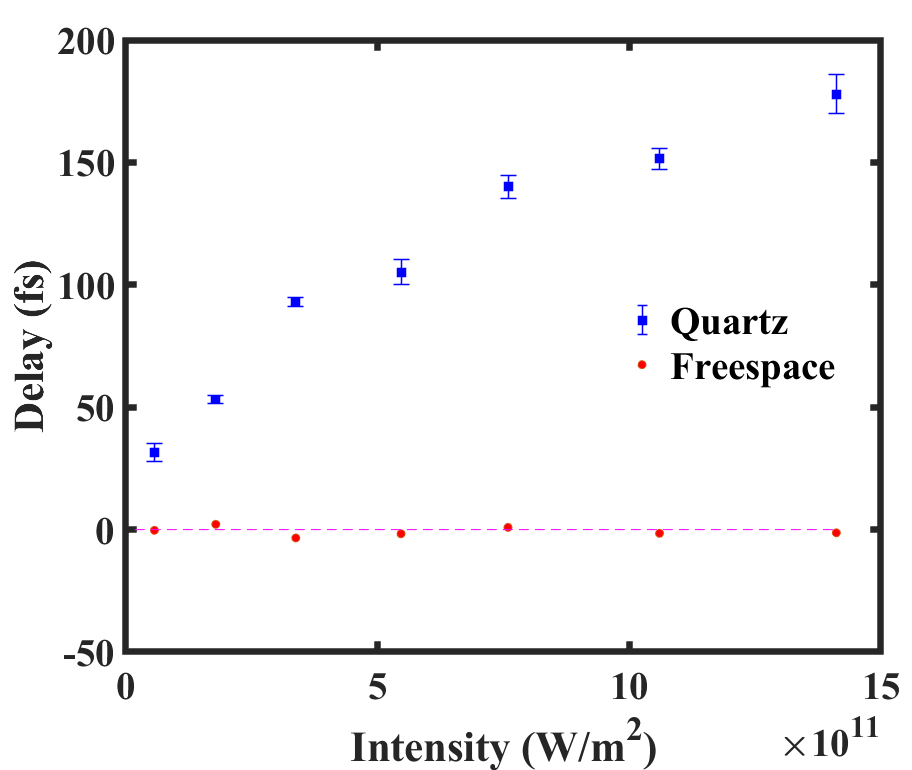} 
  \end{tabular}
\end{tabular}
\caption{\label{fig:temporal} (a) THz time-domain signal in free space (dashed lines) and crystalline quartz (solid lines) for different signal levels. The inset shows the delay increase with the growth of the THz amplitude. (b) The average time shift in free space (red) and in crystalline quartz (blue) for different signal levels. }

\end{figure*}

\section{Results and Discussion}
In Figure~\ref{fig:temporal}~(a), we show the time-domain signals for different THz field amplitudes for both free-space and crystalline quartz. One can observe an increase in the time delay experienced by the pulse in the crystal quartz with an increase of the THz field amplitude. The inset in Figure~\ref{fig:temporal}~(a) clearly demonstrates this observation. In contrast, the free-space THz time-domain signal does not exhibit such a delay increase. 

Figure~\ref{fig:temporal}~(b) shows the average time shift for each of the THz field amplitude levels compared to the lowest-level amplitude, where the average time shift for each level is calculated as
\begin{equation}
    t_{av}^i=\frac{1}{N} \sum_k^N 
t(V_i=V_k)-t(V_{low}=V_k). 
\end{equation}
Here \(t_{av}^i \ \) is the average time shift, $N$ is the number of data points, \(V_i\ \) is the $i$-th signal and \(V_{low}\ \) is the lowest-level signal. The analysis is performed over the main lobe, the interval between the first two minima, highlighted in the inset of Figure~\ref{fig:temporal}~(a),  as it represents most of the THz spectral content. We can see that, with the field intensity increase, THz pulse experiences more delay with respect to the lowest-intensity pulse. However, at higher intensities, the growth slope declines, which indicates the presence of the saturation effect. 

\begin{figure*}[t]
\centering
   \begin{tabular}{ l r }
   \begin{tabular}{@{} l @{}}
    (a) \\[1ex]
    \includegraphics[height=6cm]{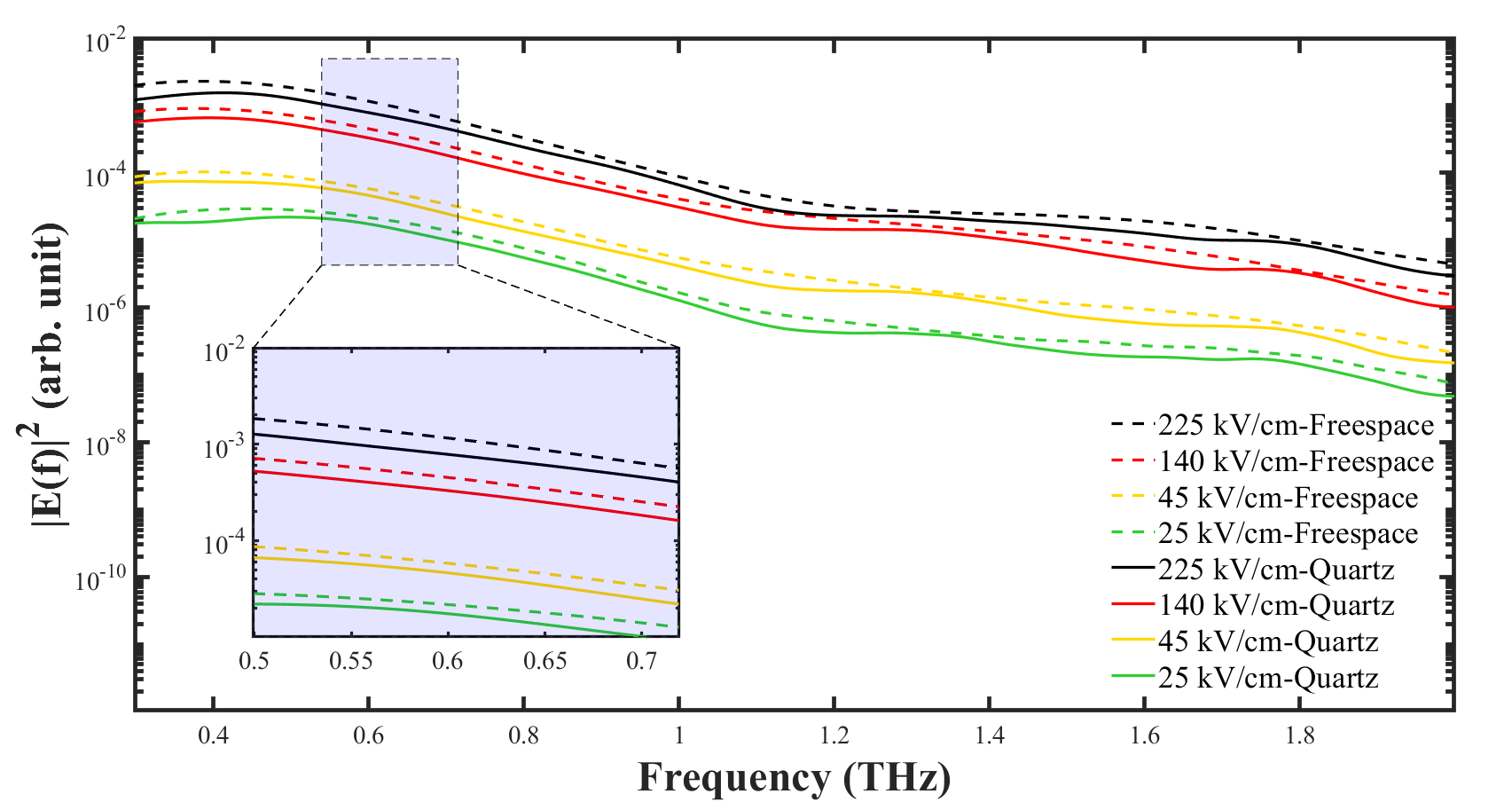}
  \end{tabular}
  &
  \begin{tabular}{l}
   (b)\\
    \includegraphics[height=5cm]{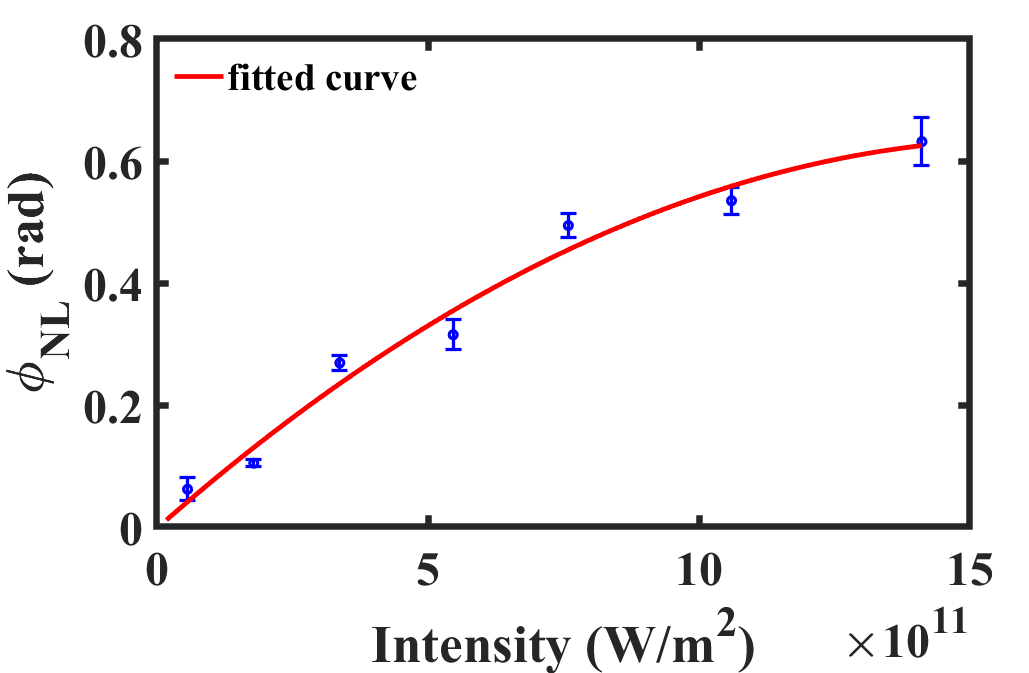} \\
    (c)\\
    \includegraphics[height=5cm]{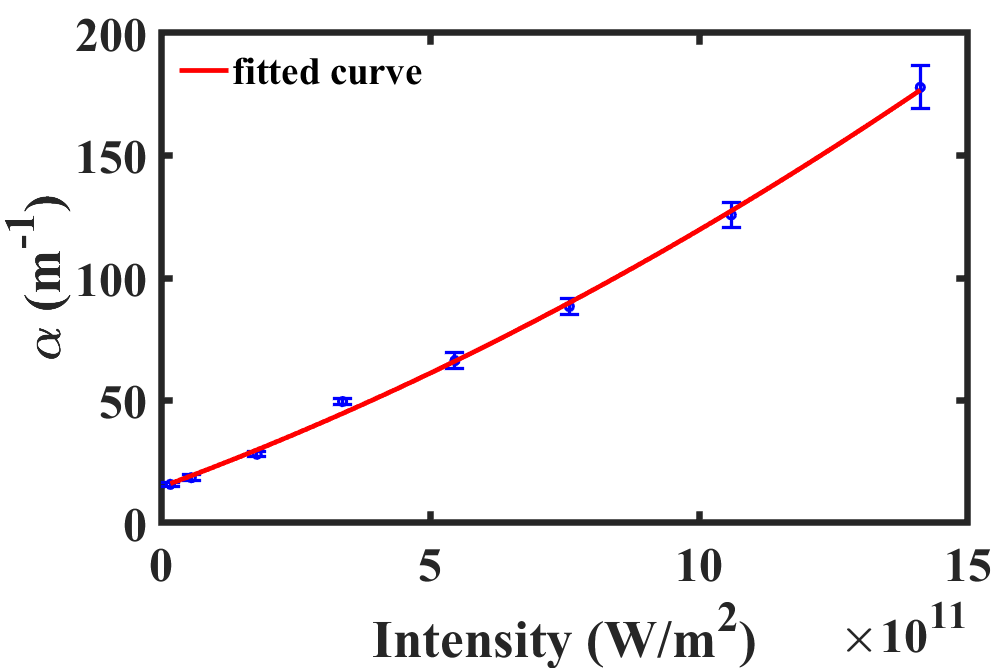} 
  \end{tabular}
\end{tabular}
\caption{\label{fig: result} (a) Signal spectral density for free-space (dashed lines) and quartz (solid lines). The difference between the free-space and quartz signals increases with an increase in the signal level. (b) nonlinear phase experienced by the signal of different amplitude at 0.4~THz. (c) absorption coefficient for each signal level at 0.4~THz.}
\end{figure*}

Figure  \ref{fig: result}~(a) shows the spectral density for the quartz sample and free space in the frequency range between 0.3 and 2~THz, where the fast Fourier transform (FFT) performed on the time-domain signal is depicted (see more details in Supporting Information). We notice that as the signal level increases, the absorption, which is the difference between the free-space and quartz sample spectra after factoring out the sample's Fresnel reflections, increases. This behavior clearly indicates the presence of a nonlinear absorption process. 

Figure \ref{fig: result}~(b) shows the nonlinear phase experienced by the THz signal for different intensity levels at 0.4~THz where the spectral density is maximum. It indicates that, as the THz intensity increases, the nonlinear phase cannot be expressed with a single linear term and suggests a negative higher-order nonlinearity term. The differential nonlinear phase for the higher THz field amplitudes in a sample with a thickness of $d$ is related to the intensity by
\begin{equation}
    \phi_i^{NL}(\omega)=\phi_i(\omega)-\phi_{low}(\omega)\ = n_2(\omega)I_i \frac{\omega}{c}d+n_4(\omega)I_i^2 \frac{\omega}{c}d.
\end{equation}
Here \(\phi_i\) is the total phase of the $i$-th signal and \(\phi_{low}\) is the phase experienced by the lowest-level signal, used as the linear response of the material, \(n_2\) and \(n_4\) are the second-order and fourth-order nonlinear refractive indices, respectively, and \(I_i\) is the peak intensity of the $i$-th level signal. By observation of Figure \ref{fig: result}~(b), one can conclude that a negative $n_4$ effect is likely to be contributing to the intensity dependence of the nonlinear phase shift.
 
 The absorption coefficient of crystalline quartz, measured as a function of the THz intensity, is depicted in Figure \ref{fig: result}~(c) (see details in Supporting Information). We can see that at lower intensities, the absorption coefficient increases linearly with respect to the field intensity. However, at higher intensities, a quadratic term also reveals itself. The absorption coefficient of the material can be expressed as
\begin{equation}
    \alpha(\omega)=\alpha_0(\omega)+\alpha_2(\omega)I_i+\alpha_4(\omega) I_i^2, 
\end{equation}
 where \(\alpha\), \(\alpha_0\), \(\alpha_2\), and \(\alpha_4\) are total absorption coefficient, linear absorption coefficient, two-photon absorption coefficient and three-photon absorption coefficient, respectively. We calculate the value of $n_2$ to be \( n_2=9.00\times 10^{-14} \ \mathrm{m^2\ W^{-1}}\). The measured value exceeds the theoretically predicted value by a factor of 1.74. This difference arises from the fact that the calculations carried out theoretically were based on the assumption that the field is monochromatic. However, the THz field used in the experiment is a short wide-band pulse. Consequently, there are contributions to the nonlinear phase shift at 0.4 THz from different frequencies, resulting in the higher value of $\phi^\mathrm{NL}$. Furthermore, the data analysis reveals the values of other nonlinear coefficients: \(n_4=-2.69\times 10^{-26} \ \mathrm{m^4\ W^{-2}}\), \( \alpha_2=8.10\times 10^{-11} \ \mathrm{ m\ W^{-1}}\) and \(\alpha_4=2.40\times 10^{-23} \ \mathrm{m^3\ W^{-2}}\). 
 The real and imaginary parts of third-order and fifth-order nonlinear susceptibility are related to these nonlinear coefficients as:
 
 \begin{subequations}
  \begin{equation}
    \label{eqchi-a}
      \Re(\chi^{(3)})= \frac{4}{3} n_0^2 \varepsilon_0 c n_2
  \end{equation}
  \begin{equation}
    \label{eqchi-b}
    \Im(\chi^{(3)})= \frac{2}{3} n_0^2 \varepsilon_0 \frac{c^2}{\omega} \alpha_2
  \end{equation}
  \begin{equation}
    \label{eqchi-c}
   \Re(\chi^{(5)})= \frac{8}{5} n_0^3 \varepsilon_0^2 c^2 n_4
  \end{equation}
  \begin{equation}
    \label{eqchi-d}
    \Im(\chi^{(5)})= \frac{4}{5} n_0^3 \varepsilon_0^2 \frac{c^3}{\omega} \alpha_4
  \end{equation}
\end{subequations}
 
 Thus, the complex third-order and fifth-order nonlinear susceptibility are found as  \( \chi^{(3)}=(1.40\times 10^{-15}+i 7.49\times 10^{-17})  \ \mathrm{m^2\ V^{-2}}\) and \( \chi^{(5)}=(-2.68\times 10^{-30}+i 1.49\times 10^{-31}) \ \mathrm{m^4\ V^{-4}}\), respectively.

\section{Conclusions}

We observe an extremely large nonlinear response of crystalline quartz in the THz region. The experimental results confirm the theoretical predictions made earlier, with an amendment to the theory by including additional vibrational resonances.  

Further, time-domain spectroscopy reveals that the observed nonlinear behavior results from a complex interplay of the third- and fifth-order susceptibilities, where the real part shows a positive third-order and a negative fifth-order contributions. Furthermore, the measured nonlinear refractive index of \( n_2=9.00\times 10^{-14} \ \mathrm{m^2\ W^{-1}}\) at 0.4~THz is seven orders of magnitude larger than the nonlinear refractive index of fused silica measured in the visible region. We attribute this large nonlinearity to the contributions from the vibrational modes in the crystal. Numerical evaluation of Equation (\ref{equ_sum}) reveals that the vibrational modes at 3.9 and 7.9 THz are the primary contributions to this large nonlinear response, despite the fact that the strong resonance at 37.2 THz might be expected to be the origin of the large optical nonlinearity.  Including these lower-frequency resonances allowed us to obtain a correct order-of-magnitude agreement between the theory and experiment. The slight difference in the values of the measured and predicted $n_2$ (a factor of 1.74)
is rooted in the fact that the spectrum of the THz radiation is wide-band, and the contribution of different spectral components is possible. This difference is a motivation for future exploration aimed at pushing the analysis beyond the approximation of a monochromatic radiation -- the necessary measure in the extremely wide-band THz frequency range.

\medskip
\textbf{Supporting Information} \par 
Supporting Information is available from the Wiley Online Library or from the authors.

\medskip
\textbf{Acknowledgements} \par 
The authors acknowledge support from Canada Research Chairs program and Natural Science and Engineering Council’s Strategic program STGP 521619. RWB acknowledges support through the Natural Sciences and Engineering Research Council of Canada, the Canada Research Chairs program, by US DARPA award W911NF-18-1-0369, US ARO award W911NF-18-1-0337, and a US Office of Naval Research MURI award N00014-20-1-2558. The authors are thankful to Prof.\ Eric VanStryland for fruitful discussion.

\medskip

%

\bibliography{template.bib}

\begin{thebibliography}{10}
\providecommand{\url}[1]{\texttt{#1}}
\providecommand{\urlprefix}{URL }

\bibitem{Peng2020}
Y.~Peng, C.~Shi, X.~Wu, Y.~Zhu, S.~Zhuang,
\newblock \emph{BME Frontiers} \textbf{2020}, \emph{2020} 2547609.

\bibitem{Lindley2021}
H.~Lindley-Hatcher, R.~I. Stantchev, X.~Chen, A.~I. Hernandez-Serrano,
  J.~Hardwicke, E.~Pickwell-MacPherson,
\newblock \emph{Applied Physics Letters} \textbf{2021}, \emph{118}, 23 230501.

\bibitem{Chen:07}
J.~Chen, Y.~Chen, H.~Zhao, G.~J. Bastiaans, X.-C. Zhang,
\newblock \emph{Opt. Express} \textbf{2007}, \emph{15}, 19 12060.

\bibitem{Jepsen2011}
P.~Jepsen, D.~Cooke, M.~Koch,
\newblock \emph{Laser \& Photonics Reviews} \textbf{2011}, \emph{5}, 1 124.

\bibitem{Kleine-Ostmann2011}
T.~Kleine-Ostmann, T.~Nagatsuma,
\newblock \emph{Journal of Infrared, Millimeter, and Terahertz Waves}
  \textbf{2011}, \emph{32}, 2 143.

\bibitem{krumbholz2009monitoring}
N.~Krumbholz, T.~Hochrein, N.~Vieweg, T.~Hasek, K.~Kretschmer, M.~Bastian,
  M.~Mikulics, M.~Koch,
\newblock \emph{Polymer Testing} \textbf{2009}, \emph{28}, 1 30.

\bibitem{Krugener2015}
K.~Kr{\"u}gener, M.~Schwerdtfeger, S.~F. Busch, A.~Soltani, E.~Castro-Camus,
  M.~Koch, W.~Vi{\"o}l,
\newblock \emph{Scientific Reports} \textbf{2015}, \emph{5}, 1 14842.

\bibitem{busch2014optical}
S.~Busch, M.~Weidenbach, M.~Fey, F.~Sch{\"a}fer, T.~Probst, M.~Koch,
\newblock \emph{Journal of Infrared, Millimeter, and Terahertz Waves}
  \textbf{2014}, \emph{35}, 12 993.

\bibitem{Nahata:96}
A.~Nahata, D.~H. Auston, T.~F. Heinz, C.~Wu,
\newblock \emph{Applied Physics Letters} \textbf{1996}, \emph{68}, 2 150.

\bibitem{Hebling:08}
J.~Hebling, K.-L. Yeh, M.~C. Hoffmann, B.~Bartal, K.~A. Nelson,
\newblock \emph{J. Opt. Soc. Am. B} \textbf{2008}, \emph{25}, 7 B6.

\bibitem{Asmontas2020}
S.~A{\v{s}}montas, S.~Bumelien{\.{e}}, J.~Gradauskas, R.~Raguotis,
  A.~Su{\v{z}}ied{\.{e}}lis,
\newblock \emph{Scientific Reports} \textbf{2020}, \emph{10}, 1 10580.

\bibitem{Hoffmann2009}
M.~C. Hoffmann, J.~Hebling, H.~Y. Hwang, K.-L. Yeh, K.~A. Nelson,
\newblock \emph{Phys. Rev. B} \textbf{2009}, \emph{79} 161201.

\bibitem{Lange2014}
C.~Lange, T.~Maag, M.~Hohenleutner, S.~Baierl, O.~Schubert, E.~R.~J. Edwards,
  D.~Bougeard, G.~Woltersdorf, R.~Huber,
\newblock \emph{Phys. Rev. Lett.} \textbf{2014}, \emph{113} 227401.

\bibitem{Tarekegne2015}
A.~T. Tarekegne, K.~Iwaszczuk, M.~Zalkovskij, A.~C. Strikwerda, P.~U. Jepsen
  \textbf{2015}, \emph{17}, 4 043002.

\bibitem{Blanchard2012}
S.~Tani, F.~m.~c. Blanchard, K.~Tanaka,
\newblock \emph{Phys. Rev. Lett.} \textbf{2012}, \emph{109} 166603.

\bibitem{Schubert2014}
O.~Schubert, M.~Hohenleutner, F.~Langer, B.~Urbanek, C.~Lange, U.~Huttner,
  D.~Golde, T.~Meier, M.~Kira, S.~W. Koch, R.~Huber,
\newblock \emph{Nature Photonics} \textbf{2014}, \emph{8}, 2 119.

\bibitem{Hafez2018}
H.~A. Hafez, S.~Kovalev, J.-C. Deinert, Z.~Mics, B.~Green, N.~Awari, M.~Chen,
  S.~Germanskiy, U.~Lehnert, J.~Teichert, Z.~Wang, K.-J. Tielrooij, Z.~Liu,
  Z.~Chen, A.~Narita, K.~M{\"u}llen, M.~Bonn, M.~Gensch, D.~Turchinovich,
\newblock \emph{Nature} \textbf{2018}, \emph{561}, 7724 507.

\bibitem{Chai2018}
X.~Chai, X.~Ropagnol, S.~M. Raeis-Zadeh, M.~Reid, S.~Safavi-Naeini, T.~Ozaki,
\newblock \emph{Phys. Rev. Lett.} \textbf{2018}, \emph{121} 143901.

\bibitem{Gaal:2006}
P.~Gaal, K.~Reimann, M.~Woerner, T.~Elsaesser, R.~Hey, K.~H. Ploog,
\newblock \emph{Phys. Rev. Lett.} \textbf{2006}, \emph{96} 187402.

\bibitem{Nelson2009}
T.~Qi, Y.-H. Shin, K.-L. Yeh, K.~A. Nelson, A.~M. Rappe,
\newblock \emph{Phys. Rev. Lett.} \textbf{2009}, \emph{102} 247603.

\bibitem{Nelson2019}
X.~Li, T.~Qiu, J.~Zhang, E.~Baldini, J.~Lu, A.~M. Rappe, K.~A. Nelson,
\newblock \emph{Science} \textbf{2019}, \emph{364}, 6445 1079.

\bibitem{Rasekh2021}
P.~Rasekh, A.~Safari, M.~Yildirim, R.~Bhardwaj, J.-M. Ménard, K.~Dolgaleva,
  R.~W. Boyd,
\newblock \emph{ACS Photonics} \textbf{2021}, \emph{8}, 6 1683.

\bibitem{Boyd2021}
A.~Tcypkin, M.~Zhukova, M.~Melnik, I.~Vorontsova, M.~Kulya, S.~Putilin,
  S.~Kozlov, S.~Choudhary, R.~W. Boyd,
\newblock \emph{Phys. Rev. Applied} \textbf{2021}, \emph{15} 054009.

\bibitem{Kerr2009}
M.~C. Hoffmann, N.~C. Brandt, H.~Y. Hwang, K.-L. Yeh, K.~A. Nelson,
\newblock \emph{Applied Physics Letters} \textbf{2009}, \emph{95}, 23 231105.

\bibitem{Tcypkin2019}
A.~N. Tcypkin, M.~V. Melnik, M.~O. Zhukova, I.~O. Vorontsova, S.~E. Putilin,
  S.~A. Kozlov, X.-C. Zhang,
\newblock \emph{Opt. Express} \textbf{2019}, \emph{27}, 8 10419.

\bibitem{Francis2020}
K.~J.~G. Francis, M.~L.~P. Chong, Y.~E, X.-C. Zhang,
\newblock \emph{Opt. Lett.} \textbf{2020}, \emph{45}, 20 5628.

\bibitem{Zalkovskij2013}
M.~Zalkovskij, A.~C. Strikwerda, K.~Iwaszczuk, A.~Popescu, D.~Savastru,
  R.~Malureanu, A.~V. Lavrinenko, P.~U. Jepsen,
\newblock \emph{Applied Physics Letters} \textbf{2013}, \emph{103}, 22 221102.

\bibitem{Dolgaleva2015}
K.~Dolgaleva, D.~V. Materikina, R.~W. Boyd, S.~A. Kozlov,
\newblock \emph{Phys. Rev. A} \textbf{2015}, \emph{92} 023809.

\bibitem{Davies2018}
C.~L. Davies, J.~B. Patel, C.~Q. Xia, L.~M. Herz, M.~B. Johnston,
\newblock \emph{Journal of Infrared, Millimeter, and Terahertz Waves}
  \textbf{2018}, \emph{39}, 12 1236.

\end{thebibliography}






  


\end{document}